%% file: resub.tex
\documentclass[aip, reprint, twocolumn, amsmath,amsfonts,amssymb]{revtex4-1}
\usepackage{threeparttable}
\input{settings}
\usepackage[colorlinks=true,linkcolor=blue]{hyperref}
\begin{document}
\title{Double Configuration Interaction Singles: Scalable and size-intensive approach for orbital relaxation in excited states and  bond-dissociation}
\author{Takashi Tsuchimochi}
\email{tsuchimochi@gmail.com}
\affiliation{College of Engineering, Shibaura Institute of Technology, 3-7-5 Toyosu, Koto-ku, Tokyo 135-8548 Japan}
\affiliation{Institute for Molecular Science, 38 Nishigonaka, Myodaiji, Okazaki 444-8585 Japan}

\begin{abstract}
We present a novel theoretical scheme for orbital relaxation in configuration interaction singles (CIS) based on a perturbative treatment of its electronic Hessian, whose analytical derivation is also established in this work. The proposed method, which can be interpreted as a ``CIS-then-CIS'' scheme, variationally accounts for orbital relaxation in excited states, thus significantly reducing the overestimation of charge-transfer excitation energies commonly associated with standard CIS. Additionally, by incorporating de-excitation effects from CIS, we demonstrate that our approach effectively describes single bond dissociation. Notably, all these improvements are achieved at a mean-field cost, with the pre-factor further reduced with the efficient algorithm introduced here, while preserving the size-intensive property of CIS. 
\end{abstract}
\maketitle
\section{Introduction}
Recently, there has been  wide spread interest in the direct determination of electronic excited states.\cite{Kowalczyk2011, Kowalczyk2013, Levi2020, Hait2021, Burton2022, Kossoski2023,Marie2023,Tuckman2023,Damour2024,Selenius2024} This surge of interest is largely due to the fact that excited states, especially those that  differ significantly from the ground state mean-field, such as core excitations\cite{Hait2020A, Garner2020, Cunha2022, arias-martinez2022} and charge transfer excitations,\cite{Liu2012,  Meyer2014, Hait2016} often favor orbital sets that   deviate considerably from Hartree-Fock (HF) or Kohn-Sham orbitals.\cite{Zhao2020, Levi2020, Hait2021, Selenius2024} As a result, excited states can be more accurately described using their own optimized orbitals, which helps mitigate the variational bias toward the ground state by allowing for significant orbital relaxation. 

Configuration Interaction Singles (CIS) is the most basic approach to studying excited states, and has been extensively investigated because of its simplicity. It is now well known that CIS tends to overestimate charge transfer excitations, for the reasons mentioned above.\cite{Subotnik2011, Liu2012} To address this issue, several studies have focused on both approximate and rigorous orbital optimizations for CIS.\cite{Liu2012, Shea2018, Shea2020, Hardikar2020}  
Liu and co-workers perturbatively estimated the orbital rotation effect by truncating the CIS energy as a quadratic function of orbital rotation parameters, which is then minimized by solving a linear equation as a Newton-Raphson step. They further approximated the second derivative of the CIS energy by the HF Hessian.\cite{Liu2012} However, it was found that the correction to charge transfer states by this treatment was generally insufficient.\cite{Liu2013} Shea and Neuscamman applied the generalized variational condition, approximately minimizing the expectation value of the square of the energy-shifted Hamiltonian within the CIS framework.\cite{Shea2018, Shea2020} They later performed full orbital optimization by requiring the CIS orbital gradient to vanish.\cite{Hardikar2020}

In this work, we take alternative, yet unexplored approach to bridging the gap between these studies. Our method is based on a rigorous CIS Hessian, which, to the best of our knowledge, we derive here for the first time. 
We then approximate it to first order to variationally account for the orbital rotation effect in CIS. Importantly, this relaxation effect remains invariant with system size, preserving a key characteristic of CIS. Interestingly, the proposed approach can also be viewed as a simple theoretical model for spontaneous emission and two-photon processes within a CIS framework, and thus can treat the ground state as well via de-excitation, offering a multi-configurational description for bond-breaking systems.

\section{Theory}
\subsection{Orbital optimization in CIS}
Throughout this paper, we use the spin orbital labels $i,j,k,l$ to represent the occupied space, $a,b,c,d$ for the virtual space, and $p,q,r,s$ for the general space, respectively. 

We consider a fully variational CIS wave function,
\begin{align}
|\tilde 0[{\bm{\kappa}},{\bf c}]\rangle = e^{-\hat \kappa} \left(\sum_{ai}c_{ai}|\Phi_i^a\rangle\right) \label{eq:SS0tilde}
\end{align}
where 
\begin{align}
    \hat \kappa &= \sum_{ai} \kappa_{ai} (\hat E_{ai}-\hat E_{ia})\label{eq:kappa}
\end{align}
and $\hat E_{ai} = a_a^\dag a_i$, and minimize its energy $E[{\bm\kappa},{\bf c}]$. The first derivatives of the energy with respect to $\kappa_{ai}$ and $c_{ai}$ around ${\bm \kappa} = {\bf 0}$ are
\begin{align}
     {^{\rm o}}g_{ai} &= \frac{\partial E[{\bm\kappa},{\bf c}]}{\partial \kappa_{ai}}\Big|_{{\bm \kappa} =  {\bf 0}} = \langle 0|\left[(\hat E_{ai} - \hat E_{ia}), (\hat H - E_0)\right]|0\rangle\\
    {^{\rm c}}g_{ai} &= \frac{\partial E[{\bm\kappa},{\bf c}]}{\partial c_{ai}}\Big|_{{\bm \kappa}  = {\bf 0}} =\langle 0 |(\hat H -E_0) |\Phi_i^a\rangle
    +\langle \Phi_i^a |(\hat H -E_0) |0\rangle
\end{align}
where $E_0$ is the energy evaluated at ${\bm \kappa} = {\bf 0}$. While ${^{\rm c}}g_{ai}$ represent the CIS residuals, ${^{\rm o}}g_{ai}$ correspond to the Fock-like matrix elements in the sense of the Brillouin theorem extended to CIS. We use the superscripts ``o'' and ``c'' to distinguish between the orbital rotation contribution and the CI contribution. Below, the CI-related terms are often labeled with the Greek letters $\mu,\nu$. 

A CIS state is stationary when the first derivatives of the energy with respect to both orbital rotations and CI coefficients vanish, i.e., when both the orbital rotation gradients (${^{\rm o}}{\bf g}$) and CI gradients (${^{\rm c}}{\bf g}$) are zero. To achieve this stationary condition, one approach is to optimize the orbitals and the CI coefficients separately. This method, which decouples the two optimization processes, was previously explored in Ref.[\onlinecite{Hardikar2020}].

However, it is well known that such a decoupled approach can lead to slow convergence, since the orbital rotation and CI problems are inherently coupled, as in CASSCF.\cite{Helgaker2000} To address this issue, it is desirable to introduce the electronic Hessian, which for real orbitals can be expressed as
\begin{align}
      {^{\rm oo}}H_{ai,bj}  &= {\mathscr P}(ai){\mathscr P}(bj) (2\;{^{\rm oo}}A_{ai,bj} +   {^{\rm oo}}B_{ai,bj} + {^{\rm oo}}B_{bj,ai})\\
    {^{\rm oc}}H_{ai,\nu}    &= 2{\mathscr P} (ai) (  {^{\rm oc}}A_{ia,\mu} +  {^{\rm oc}}B_{ia,\mu} - {^{\rm o}}g_{ai} c_\nu)\\
       {^{\rm cc}}H_{\mu\nu} &=2\; {^{\rm cc}} A_{\mu\nu}
   -{^{\rm c}}g_{\mu} c_\nu
   -{^{\rm c}}g_{\nu} c_\mu
\end{align}
where ${\mathscr P}(pq)$ is the anti-symmetrizer between $p$ and $q$, and 
\begin{align}
{^{\rm oo}}A_{pq,rs}&=\langle 0|\hat E_{qp} (H-E_0) \hat E_{rs}|0\rangle\\
{^{\rm oc}}A_{pq,\mu}&=\langle 0|\hat E_{qp} (\hat H - E_0) |\Phi_\mu\rangle\\
{^{\rm cc}} A_{\mu\nu}&=\langle \Phi_\mu| (\hat H - E_0) |\Phi_\nu\rangle \\
{^{\rm oo}}B_{pq,rs}&=\langle 0|(H-E_0) \hat E_{pq} \hat E_{rs}|0\rangle\\
{^{\rm oc}}B_{pq,\mu}&=\langle 0| (\hat H - E_0) \hat E_{pq}|\Phi_\mu\rangle
\end{align}
Note that ${^{\rm oo}}A_{ai,bj}$ and ${^{\rm oo}}B_{ai,bj}$ resemble the familiar matrices that appear in the Hartree-Fock electronic Hessian. However, unlike in Hartree-Fock, the de-excitation terms such as ${^{\rm oo}}A_{ia,jb}$ and  ${^{\rm oo}}B_{ia,jb}$ are also non-zero.

The explicit working equations for evaluating these terms are provided in the Supplementary Material.\cite{si}
 
\subsection{Double CIS}
Even with the derived Hessian, orbital optimization for excited states can be challenging. Since the Hessian is never positive definite, straightforward treatments of this highly non-linear problem may be unstable especially for high-lying excited states. This can lead to convergence to other lower excited states instead of the desired state, necessitating careful manipulation  to avoid the so-called variational collapse.\cite{Gilbert2008, Hait2020B, CarterFenk2020, Hardikar2020}
 
To ameliorate this complexity, it is desirable to make the optimization process black-box and ensure guaranteed convergence. To this end, our starting point is to recognize that, to first order, orbital rotation corresponds to a set of single excitations, including the reference state, i.e., $e^{-\hat\kappa}\approx 1 - \hat\kappa$. Therefore, if the target CIS state is close to the stationary state ($\hat \kappa\approx 0$), it is expected that the orbital relaxation can be well approximated by performing a simple CIS on top of the original CIS --- hence dubbed Double CIS (DCIS). 

Since CIS is invariant with respect to rotations among the occupied and virtual spaces, the parameterization for $\hat \kappa$ as given in Eq.(\ref{eq:kappa}) is sufficient. Hence, an optimized state may be given by
\begin{align}
    |\Psi\rangle &= (1-\hat\kappa)|0\rangle \br
    &=|0\rangle -\sum_{aibj}\kappa_{ai}c_{bj}|\Phi_{ij}^{ab}\rangle 
    +(\sum_{ai} \kappa_{ai}c_{ai})|\Phi_0\rangle \label{eq:Psi}
\end{align}
This form is indeed identical to that explored previously by Liu and Subotnik\cite{Liu2012}, called orbital-optimized CIS (OOCIS), where $\kappa_{ai}$ were approximated perturbatively with the Hartree-Fock Hessian ${^{\rm oo}} {\bf H}^{\rm HF}$. Specifically, they considered the following quadratic energy function,
\begin{align}
E_{\rm OOCIS}[{\bm \kappa}] = E_0 + \sum_{ai} {^{\rm o}} g_{ai} \kappa_{ai} + \frac{1}{2} \sum_{abij} {^{\rm oo}} H^{\rm HF}_{ai,bj} \kappa_{ai}\kappa_{bj} 
\end{align}
and minimized it by solving the linear equation
\begin{align}
    \sum_{bj} {^{\rm oo}} H^{\rm HF}_{ai,bj} \kappa_{bj} + {^{\rm o}}g_{ai} = 0
\end{align}
They further extended their scheme to generate several perturbed CIS states and constructed an effective Hamiltonian matrix in this subspace, which was then diagonalized; thus falling into the category of the ``perturb-then-diagonalize'' approach.\cite{Liu2013, Liu2014} 

In contrast, we aim to ``variationally'' determine the orbital optimized CIS. To ensure the rigor of our derivation, it is important to note that Eq.(\ref{eq:Psi}) accounts only for orbital rotation and neglects the first-order response of the CI coefficients, i.e., the coupling effects between them. Therefore, it should be emphasized that the perturbative treatment is incomplete. 

To include this response in a computationally efficient and stable manner, we can treat the perturbative correction $|\bar {\bf c}\rangle$ to the CI coefficients of $|0\rangle$ as follows:\cite{Helgaker2000} 
\begin{align}
    |\Psi_{\rm DCIS}\rangle = (1-\hat \kappa)|0\rangle + \left(1-|0\rangle \langle 0|\right) |\bar {\bf c}\rangle \label{eq:DCIS}
\end{align}
Here, $|\bar {\bf c}\rangle = \bar c_0 |\Phi_0\rangle + \sum_{ai}\bar c_{ai}|\Phi_i^a\rangle$ represents a generalized CIS ansatz.\cite{Shea2018} 
We intentionally introduced the HF determinant $|\Phi_0\rangle$, to account for the response in the ground state and its orthogonality. 
Eq.(\ref{eq:DCIS}) clearly encompasses the first order corrections to both orbital rotation and CI coefficients in terms of Eq.(\ref{eq:SS0tilde}). It should be emphasized that our method differs from the OOCIS scheme by having a well-defined wave function (\ref{eq:DCIS}), whose energy is given as the expectation value of the Hamiltonian, whereas there is no associated wave function in OOCIS.

It is more convenient to {\it equivalently} rearrange the variational parameters in Eq.(\ref{eq:DCIS}), without loss of generality, so that the DCIS state is expressed as
\begin{align}
     |\Psi_{\rm DCIS}\rangle 
     &= \sum_{ai}d_{ai}\hat E_{ai}|0\rangle + \sum_\mu \bar d_\mu |\Phi_\mu\rangle
     \label{eq:DCIS_new}
\end{align}
which is regarded as CIS on top of CIS. In fact, it can be easily shown\cite{si} that its variational space is identical to that of the ``internally-contracted'' multi-reference CIS with the CIS reference, i.e., $\sum_{pq} t_{pq}\hat E_{pq} |0\rangle$. While the internally-contracted form may be more appealing because of its clear physical interpretation, the form presented in Eq.(\ref{eq:DCIS_new}) is far preferred in practice for two reasons: (1) the internally-contracted CIS can introduce a significant number of redundancies, posing considerable challenges in optimization whereas Eq.(\ref{eq:DCIS_new}) is free of such redundancies, and (2) it allows for a direct translation of the results from the electronic Hessian into the DCIS working equations, requiring minimal modification to the Hessian program, as shown below.

With the DCIS variational ansatz defined, we can derive the set of linear equations needed to minimize the energy expectation value $E_{\rm DCIS} = \langle \Psi_{\rm DCIS}|\hat H |\Psi_{\rm DCIS}\rangle$ under the normalization condition:
\begin{subequations}
\begin{align}
 \langle 0|\hat E_{kc} (\hat H - E_{\rm DCIS}) |\Psi_{\rm DCIS}\rangle &= 0\\
 \langle \Phi_\mu| (\hat H - E_{\rm DCIS}) |\Psi_{\rm DCIS}\rangle &= 0\label{eq:eq}
\end{align}
\end{subequations}
Interestingly, the DCIS Hamiltonian matrix, the second derivative of the energy expectation value with respect to $\{d_{ai}, \bar d_0, \bar d_{ai}\}$, parallels the electronic Hessian matrix {\bf H}. 
Namely, these equations are recast as the DCIS generalized eigenvalue problem:
\begin{align}
    \begin{pmatrix}
        {^{\rm oo}}{\bf A} & {^{\rm oc}}{\bf A}\\
        {^{\rm co}}{\bf A} & {^{\rm cc}}{\bf A}
    \end{pmatrix}
    \begin{pmatrix}
        {\bf d}\\ \bar {\bf d}
    \end{pmatrix}
    = \omega    \begin{pmatrix}
        {^{\rm oo}}{\bf S} & {^{\rm oc}}{\bf S}\\
        {^{\rm co}}{\bf S} & {^{\rm cc}}{\bf S}
    \end{pmatrix}
    \begin{pmatrix}
        {\bf d}\\ \bar {\bf d}
    \end{pmatrix}
\end{align}
where $\omega = E_{\rm DCIS} - E_{0,I}$ represents the relaxation energy of the $I$th state, and ${\bf S}$ is the overlap matrix,
\begin{align}
{^{\rm oo}}S_{ai,bj}&=\langle 0|\hat E_{ia} \hat E_{bj}|0\rangle\\
{^{\rm oc}}S_{ai,\mu}&=\langle 0|E_{ia} |\Phi_\mu\rangle\\
{^{\rm cc}} S_{\mu\nu}&=\delta_{\mu\nu}\end{align}

\subsubsection{Properties of DCIS}
There are several important aspects about DCIS that we should clarify.

{\it Size-intensivity}. The excitation energy of DCIS is essentially size-{\it intensive}. In other words, for a non-interacting system AB where A is sufficiently distant from B, the relaxation energy of AB is the same as that computed solely for the system A (or B). However, a caveat arises when the target CIS state is exactly degenerate, like all other state-specific methods. Namely, if A and B are the same molecule having the identical geometry, any linear combination between two ``local'' CIS states of A and B can form a whole CIS state of AB, which is yet not a product state, i.e., $|\Psi_{\rm CIS}^{\rm AB}\rangle=C_A|\Psi_{\rm CIS}^{\rm A}\rangle|\Phi_0^{\rm B}\rangle + C_B|\Phi_0^{\rm A}\rangle|\Psi_{\rm CIS}^{\rm B}\rangle \ne |\Psi_{\rm CIS}^{\rm A}\rangle|\Psi_{\rm CIS}^{\rm B}\rangle$. This causes an inconsistency where the DCIS energy is not invariant with respect to rotation between degenerate CIS states. This issue occurs because DCIS is a state-specific, linear method. 
Fortunately, this is a rare event and we do not expect this to happen in practice. Furthermore, one can easily circumvent this issue by extending DCIS to a multi-state scheme similar to Ref.[\onlinecite{Liu2013}], although such an extension is beyond the scope of this work. 

{\it Bond dissociation}. DCIS can, in principle, describe single bond-breaking as it includes doubly excited configurations $|\Phi_{ij}^{ab}\rangle$ as well as the HF determinant $|\Phi_0\rangle$. By diagonalizing the DCIS Hamiltonian, one not only obtains the energy relaxation contribution to the target CIS state, but also accesses other variational states. Of particular interest is the lowest energy solution, which is guaranteed to have a negative excitation energy representing ``de-excitation'' and a total energy lower than that of the HF state, thus representing the ground state with a large weight $\bar d_0$. 
The ground state also benefits from approximate orbital rotation effects, thanks to the presence of singles. It is noteworthy that the coefficients for doubles are constrained since $c_{ai}$ are fixed, and hence its performance must be tested.

{\it Double excitation}. Similarly to the previous point, the presence of $|\Phi_{ij}^{ab}\rangle$ may allow us to determine excited states with significant double excitation character, approximated by the amplitude $c_{ai}d_{bj}$. This is a unique feature about DCIS, mimicking the two-photon absorption process. 

{\it Spin adaptation}. By employing spin-adapted singles, spin-contamination can be completely avoided. It is also straightforward to obtain triplet states in place of singlets.

{\it Computational scaling}. The formal scaling of  DCIS remains at $O(N^4)$ where $N$ is the system size, identical to that of CIS. However, computing the gradients ${^{\rm o}}{\bf g}$ requires the formation of two Fock-like matrices.\cite{si} Although this is only necessary once, calculating each of the sigma-vectors with the trial vectors $^{\rm o}{\bf x}$ and $^{\rm c}{\bf x}$ involves forming five Fock-like matrices to evaluate ${^{\rm oo}}{\bf A}{^{\rm o}}{\bf x}$, ${^{\rm oc}}{\bf A}{^{\rm c}}{\bf x}$,  ${^{\rm co}}{\bf A}{^{\rm o}}{\bf x}$, and ${^{\rm cc}}{\bf A}{^{\rm c}}{\bf x}$.\cite{si} As a result, DCIS is approximately five times more expensive than CIS per cycle. However, this disadvantage is largely offset by the fact that CIS requires determining basically all the low-lying $(I-1)$ states, whereas DCIS can bypass this computational burden, as discussed below.

\subsubsection{Efficient computational scheme}\label{sec:algorithm}
A clear advantage of DCIS lies in the fact that its solution is guaranteed to be obtained stably, in contrast to the highly non-linear orbital optimization process. Typically, the target DCIS solution appears as the $(I+1)$th {\it excited state} of the DCIS Hamiltonian {\bf A}; other low-lying $I$ DCIS states, including the ground state, are simultaneously obtained with smaller (negatively larger) eigenvalues. This can turn to a downside of the method, as it could become five times more demanding than CIS, if the standard Davidson algorithm\cite{Davidson1975, Sadkane1999} is employed to diagonalize {\bf A}, requiring many cycles to reach the $(I+1)$th solution. 

However, given that the target DCIS state $|\Psi_{\rm DCIS}\rangle$ is expected to represent a small (perturbative) correction to the CIS state $|0_I\rangle$, we can ``follow'' the target state during iterative optimization with the Davidson algorithm, bypassing the other low-lying eigenstates. Specifically, we initialize with $|0_I\rangle$ (i.e., $\bar d_{ai} = c_{ai}^I$ and $\bar d_0 = d_{ai} = 0$) and compute the overlap between the CIS and trial DCIS states, $|\langle 0_I | \Psi^{(k)}_{K, \rm DCIS} \rangle|$, at each $k$th cycle, to identify the corresponding trial vector $|\Psi^{(k)}_{I, \rm DCIS} \rangle$ with the largest overlap. We then form the perturbative update {\it only} for $|\Psi^{(k)}_{I, \rm DCIS} \rangle$ by using diagonal preconditioning, as opposed to the block Davidson algorithm,\cite{Sadkane1999} which updates multiple states simultaneously and therefore requires to calculate Fock-like matrices many times. While alternative methods like DIIS\cite{Pulay1980,Pulay1982,Hardikar2020} could be employed, the proposed "maximum-overlap" algorithm guarantees orthogonality from other low-lying states and is stable, with a simple modification to the Davidson algorithm. The pseudocode is presented in the Supplementary Material.\cite{si}

We note that the maximum-overlap Davidson algorithm does not suffer from the root-flipping issue commonly encountered in its SCF variants.\cite{Gilbert2008, Hait2020B, CarterFenk2020} Even if the target trial state $|\Psi^{(k)}_{I, \rm DCIS} \rangle$ exhibits some degeneracy, such that the overlaps $|\langle 0_I | \Psi^{(k)}_{I, \rm DCIS} \rangle|$ and $|\langle 0_I | \Psi^{(k)}_{J\ne I, \rm DCIS} \rangle|$ accidentally become identical throughout the calculation, convergence is still assured. This is because, at each iteration, a new trial state orthogonal to the current subspace is always added, ensuring that the Krylov subspace continues to expand and eventually spans the entire DCIS Hilbert space. In such cases, the algorithm may converge more slowly, potentially requiring more Fock builds than the block Davidson algorithm.
However, we have not encountered this issue in practice. As demonstrated below, the maximum-overlap algorithm achieves significant speed-ups compared to the block Davidson algorithm.

\begin{table}[b!]
    \caption{Size-intensivity test for HF$\cdots$(Be)$_n$ (eV).}
    \label{tab:size_intensivity}
    \tabcolsep = 0.5em
    \begin{tabular}{c|rrrr}
     \hline\hline
     State & $n=0$ & $n=1$& $n=2$  \\
     \hline
      CIS ($\pi-\sigma^*$)  & 10.6163 & 10.6165 & 10.6165 \\
      DCIS ($\pi-\sigma^*$) & -2.4768 & -2.4770 & -2.4770\\
      DCIS (Ground State) &-10.7013 &   -10.7015 &   -10.7015 \\
    \hline\hline
    \end{tabular}
\end{table}

\begin{figure*}
    \includegraphics[width=1\linewidth]{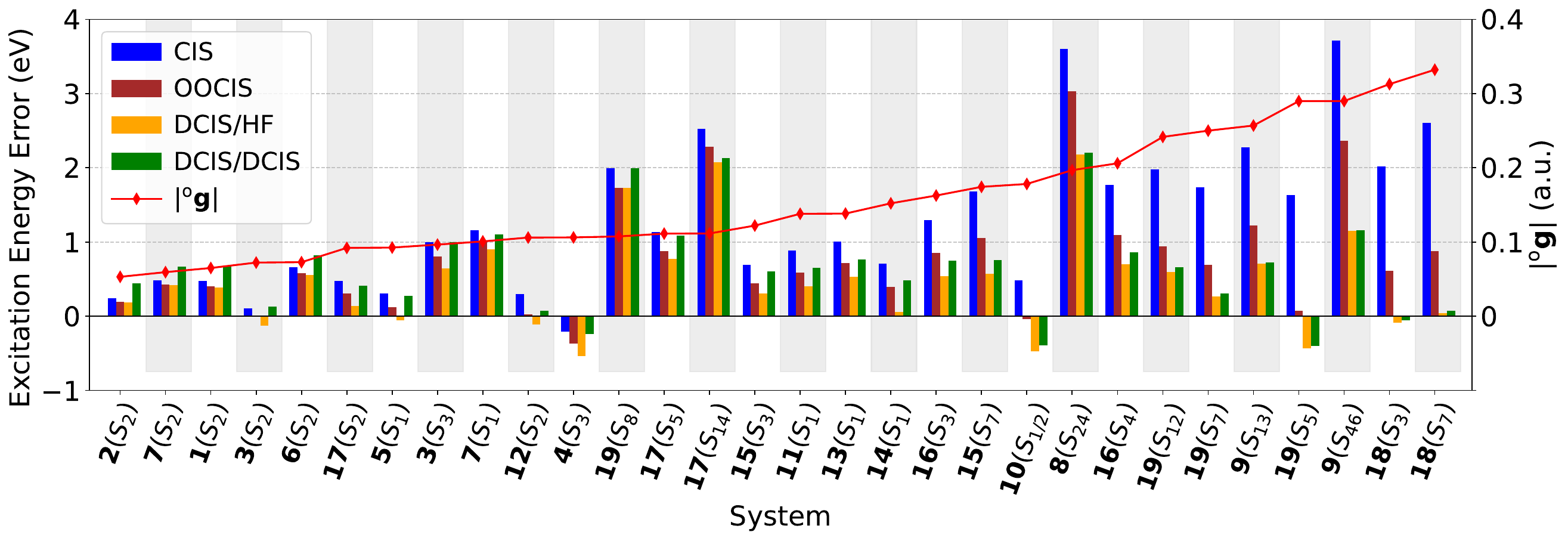}
    \caption{Errors from the theoretical best estimates (with cc-pVTZ) for charge transfer excitations of 19 structures, taken from Ref.\onlinecite{Loos2021}. DCIS/HF and DCIS/DCIS indicate the choice of ground state (HF or DCIS). The systems are ordered by the magnitude of the CIS orbital gradient, and the state numbers $S_n$ are those of CIS.}
    \label{fig:CT}
\end{figure*}

\section{Results}

We have implemented both the CIS Hessian and DCIS in our in-house code. However, it is not our intention to investigate the performance of Hessian-based orbital optimization of CIS; its detailed investigation will be conducted in future studies. Instead, we focus here on DCIS and demonstrate its capabilities. To do so, we first examine its size-intensive property, and present results for charge transfer excitations and bond dissociation. 
 We will then demonstrate that the maximum-overlap algorithm considerably reduces the computational cost compared to the standard Davidson algorithm.

\subsection{Size-intensivity test}
As direct numerical evidence of the size-intensivity of DCIS, we calculated the DCIS relaxation energy of a CIS excitation using the HF molecule with $n$ Be atoms separated by 10 {\AA}. Table~\ref{tab:size_intensivity} shows the CIS excitation energy of $\pi-\sigma^*$ transition computed at a bond length of $1.0$ {\AA} using a 6-31G basis set for $n=0,1,2$, along with the DCIS relaxation effect on this CIS state. Clearly, the results confirm that DCIS is size-intensive, as the relaxation remains unchanged with system size. Additionally, the DCIS ground state, as the lowest root, exhibits the same invariance character, demonstrating that the correlation energy is size-intensive, though not size-consistent, similar to spin-flip (SF) CIS.\cite{Krylov2001}

\subsection{Charge transfer excitations}\label{sec:CT}
For charge transfer excitations, the orbital derivative $^{\rm o}{\bf g}$ is typically large, indicating that a significant orbital relaxation effect is needed to lower the energies and mitigate the HF orbital bias. Since DCIS incorporates this relaxation in CIS, it is expected to improve excitation energies, particularly compared to OOCIS, because the latter approximates the CIS Hessian by that of HF.\cite{Liu2012}

\begin{figure}
    \includegraphics[width=0.8\linewidth]{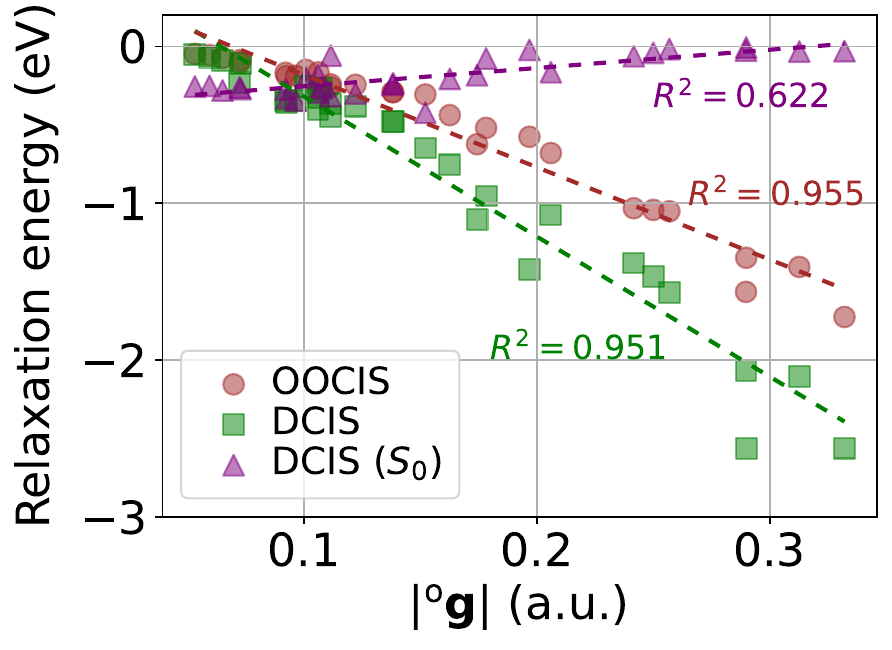}
    \caption{Relaxation energy obtained by OOCIS (brown), and by diagonalization of the DCIS Hamiltonian for ground (purple) and excited (green) states, as a function of the CIS orbital gradient.}
    \label{fig:relax}
\end{figure}
To demonstrate this, we computed the charge transfer excitation energies for 30 states across 17 organic compounds (19 structures), using the data set from Loos et al..\cite{Loos2021} These results were compared with the theoretical best estimates based on coupled cluster methods with cc-pVTZ.\cite{Loos2021} Fig.\ref{fig:CT} summarizes the errors in excitation energies for these charge transfer states, ordered in ascending order of the CIS gradient $|^{\rm o}{\bf g}|$ (red line). As expected, the CIS error generally increases with the magnitude of $|^{\rm o}{\bf g}|$, while DCIS significantly improves the excitation energies. 
Note that there are two reference ground state options for the DCIS results. If the HF state is used as the ground state, as in OOCIS, then the results directly reflect the relaxation of the excited state. Alternatively, one can use the DCIS lowest eigenstate, which preserves orthogonality and facilitates straightforward evaluations of inter-state couplings such as transition dipole moments. While the former choice (DCIS/HF) usually outperforms the latter (DCIS/DCIS), the states inevitably become nonorthogonal.

OOCIS also improves excitation energies, but to a lesser extent than DCIS, particularly for systems with considerable $|^{\rm o}{\bf g}|$, such as {\bf 9} ($\beta$-dipeptide), {\bf 18} (twisted-dimethylaminobenzonitrile), and {\bf 19} (twisted N-phenylpyrrole). This may be partly attributed to OOCIS being a one-shot perturbative correction. More importantly, we believe that the use of the HF orbital Hessian in place of that of CIS may not be adequate. After all, the HF orbitals themselves are not reliable for these states; then, why should the HF Hessian be?

Fig.~\ref{fig:relax} presents the relaxation energy provided by OOCIS and DCIS for each excited state as a function of the CIS gradient. As expected, the OOCIS relaxation effect is consistently weaker than that of DCIS, which rigorously employs part of the CIS Hessian information ({\bf A}). Both methods yield similar results for states with $|^{\rm o}{\bf g}|<0.15$ but the difference becomes more pronounced with increasing $|^{\rm o}{\bf g}|$. 
In passing, the DCIS ground state correlation energy is sufficiently small across all systems as shown in Fig.~\ref{fig:relax}, further supporting its size-intensivity. However, the magnitude of the ground state correlation energy is typically falls within the range of 0.1-0.2 eV, and therefore the DCIS/DCIS results are slightly overestimated compared to those of DCIS/HF. 

While the orbital gradient of CIS energy has a fairly strong correlation with the relaxation energy as shown in Fig. \ref{fig:relax} ($R^2=0.951$ and $0.955$ for DCIS and OOCIS, respectively), it does not provide direct insight into how much these relaxation energies actually improve the description of excitations involving spatially significant charge transfer. That is to say, the large $|^{\rm o}{\bf g}|$ does not always mean strong charge transfer.

\begin{figure}
    \centering
    \includegraphics[width=0.7\linewidth]{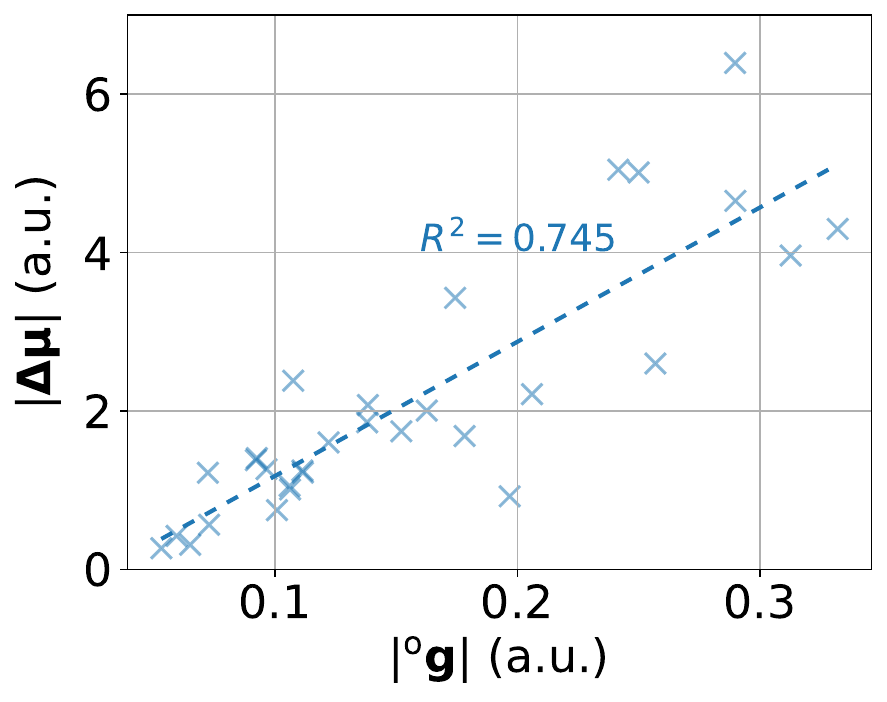}
    \caption{Correlation between CIS orbital gradient ($|^{\rm o}{\bf g}|$) and dipole change ($|{\bm\Delta}{\bm\mu}|$) using the same test set as Fig. \ref{fig:CT}.}
    \label{fig:g_mu}
\end{figure}

The extent of charge transfer is often characterized by the difference between the dipole moments of the ground and excited states, $|{\bm\Delta}{\bm\mu}|$. In Fig. \ref{fig:g_mu}, we found that $|^{\rm o}{\bf g}|$ is also linearly correlated with $|{\bm\Delta}{\bm\mu}|$ with $R^2 = 0.745$, 
though the correlation is weaker than with the relaxation energy. This motivates further investigation into 
whether DCIS provides a better mean-field description of excited states 
with significant charge transfer, using $|{\bm\Delta}{\bm\mu}|$ as a metric. Such a test should also offer more detailed insight into whether DCIS and OOCIS can be useful for charge transfer states.

\begin{figure}
    \includegraphics[width=0.8\linewidth]{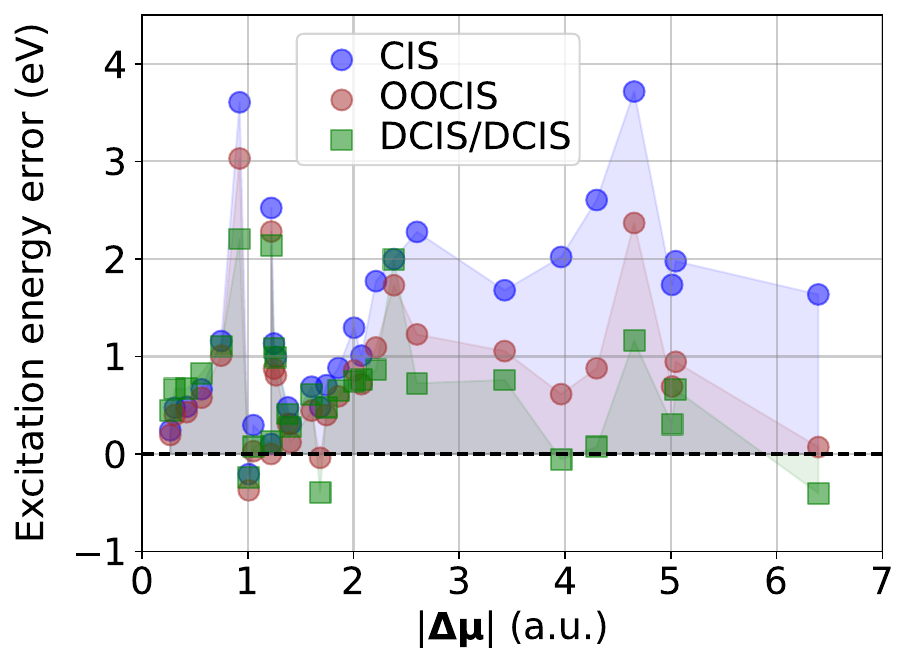}
    \caption{Error in excitation energy computed with CIS (blue), OOCIS (brown), and DCIS/DCIS (green), as a function of the dipole change. The error area is filled with color for visual guidance.}
    \label{fig:dipole}
\end{figure}

To that end, in Fig. \ref{fig:dipole}, we plotted the same excitation energy errors shown in Fig. \ref{fig:CT} but now as a function of $|{\bm\Delta}{\bm\mu}|$, as computed by HF and CIS. The results in Fig. \ref{fig:dipole} clearly demonstrate that DCIS consistently improves the excitation energy for large $|{\bm\Delta}{\bm\mu}|$ (greater than about 2.5 {\it a.u.}) more significantly than OOCIS does. On the other hand, as expected, there is little to no improvement with DCIS or OOCIS over CIS when $|{\bm\Delta}{\bm\mu}|$ is smaller, with the exception of the dipeptide ({\bf 8}). Although $|{\bm\Delta}{\bm\mu}|$ is only $0.92$ {\it a.u.} for this CIS state, which is arguably too small to be classified as charge transfer, we observed a notable reduction in error with DCIS, by 1.4 eV, due to the large CIS orbital gradient, $|^{\rm o}{\bf g}| \approx 0.2$ {\it a.u.}. 
This large orbital rotation effect and error reduction are thus attributed not to charge transfer, but to the state simply being high-lying, as it corresponds to the 24th CIS state.

We have summarized the mean-errors (MEs) and mean-absolute-errors (MAEs) of each method in Table \ref{tab:MAE}. Our results indicate that DCIS/HF maintains similar accuracy regardless of the excitation character, offering a balanced description. As discussed above, DCIS/DCIS is slightly less accurate than DCIS/HF, mainly due to the small amount of ground state correlation introduced, which leads to an overestimation of valence excitations (systems with $|{\bm\Delta}{\bm \mu}|<0.25$). Nonetheless, both DCIS/HF and DCIS/DCIS significantly improve the description of charge transfer excitations ($|{\bm\Delta}{\bm \mu}|>0.25$), achieving MAEs of 0.48 and 0.52 eV, respectively. In stark contrast, OOCIS actually becomes less accurate for charge transfer states, with an MAE of 0.98 eV, which is approximately 0.3 eV worse than for valence states.

Overall, while OOCIS reduces the errors in CIS charge transfer states by half, DCIS achieves an additional reduction of the remaining error by half. However, it is important to note that the orbital rotation effect in DCIS is still an approximation of the full orbital rotation. Further studies are necessary to evaluate the significance of higher-order terms present in the full orbital rotation.

\begin{table}[t!]
    \caption{ME and MAE of each method in eV.}
    \label{tab:MAE}
    \begin{threeparttable}
    \begin{tabular}{c|rrrrr}
    \hline\hline
         & CIS & OOCIS & DCIS/HF & DCIS/DCIS \\
         \hline
         ME (all)& 1.29 & 0.78 & 0.47 & 0.66\\
         MAE (all) & 1.30 & 0.80 & 0.59 & 0.73\\
         %ME (CT)$^{\rm a}$ & 1.98  & 1.01 & 0.43 & 0.52  \\
         %MAE (CT)$^{\rm a}$& 1.98 & 1.01 & 0.59 & 0.66 \\
         ME (Val)$^{\rm a}$ & 0.68 & 0.64 & 0.55 & 0.83\\
        MAE (Val)$^{\rm a}$  & 0.70 & 0.69 & 0.66 & 0.87 \\        
        ME (CT)$^{\rm b}$ & 2.21 &
        0.98  & 0.35 &  0.40 \\
          MAE (CT)$^{\rm b}$ &  2.21 &  0.98  & 0.48 &  0.52 \\
         \hline\hline
    \end{tabular}
        \begin{tablenotes}
        \small
          \item[a] Systems with $|{\bm\Delta}{\bm\mu}|<2.5$.
          \item[b] Systems with $|{\bm\Delta}{\bm\mu}|>2.5$.
    \end{tablenotes}
\end{threeparttable}
\end{table}

\begin{figure}[t!]
    \centering
    \includegraphics[width=0.99\linewidth]{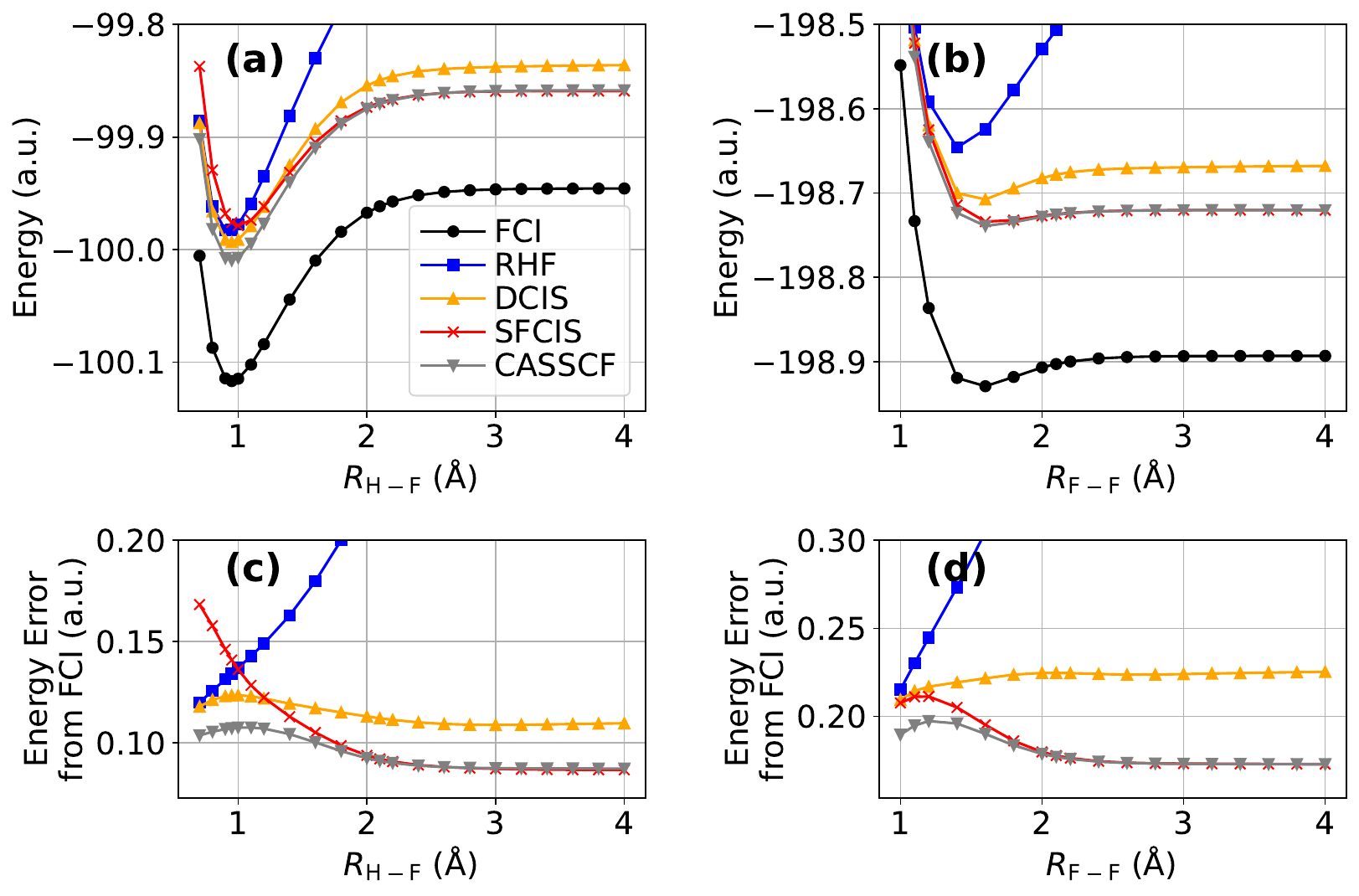}
    \caption{Potential energy curves for (a) HF and (b) F$_2$. Errors from FCI are depicted in (c) and (d).}
    \label{fig:DCIS_PES}
\end{figure}

\begin{table}[b]
    \centering
    \caption{Non-parallelity error (NPE) in mH, defined as the difference between the maximum and minimum errors from the FCI potential cruve.}
    \begin{tabular}{c|rrrrr}
    \hline
    \hline
         &   HF & DCIS & SFCIS & CASSCF(2e,2o)\\
\hline
         HF& 247.1 & 14.7 & 81.7 & 20.3\\
         F$_2$& 342.8 & 15.9 & 38.4 & 24.6\\
\hline\hline
    \end{tabular}
    \label{tab:NPE}
\end{table}

\begin{figure*}
    \centering
    \includegraphics[width=0.8\linewidth]{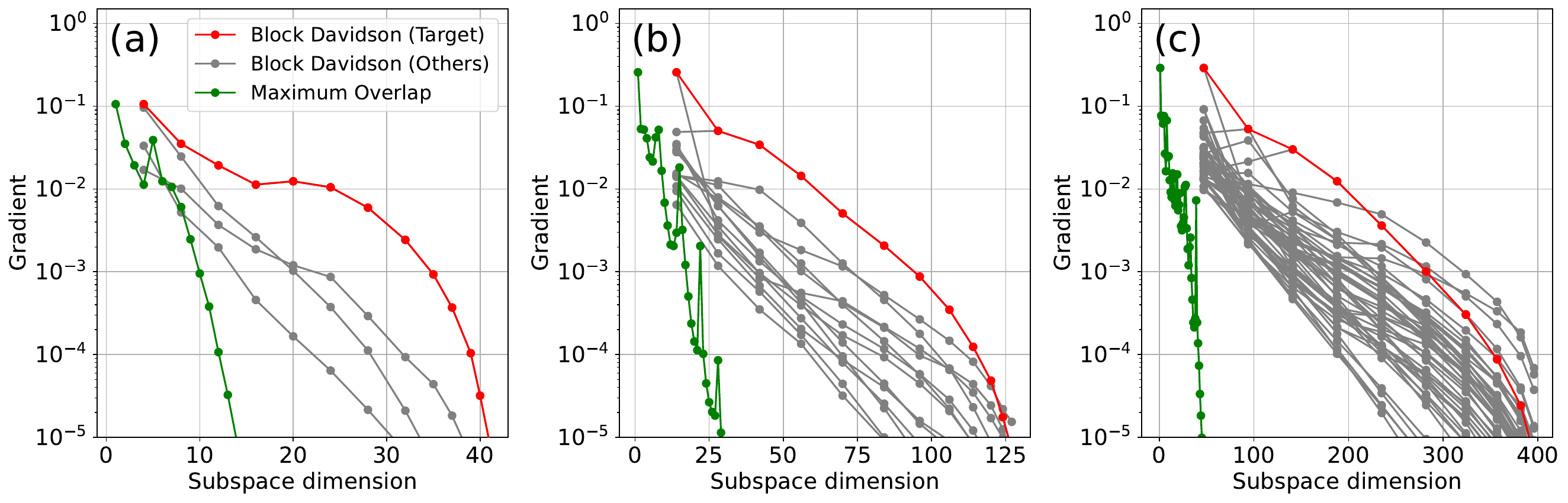}
    \caption{Convergence behavior of block Davidson and maximum-overlap Davidson algorithms. (a) $S_3$ of benzonitrile ({\bf 4}). (b) $S_{13}$ of $\beta$-Dipeptide  ({\bf 9}). (c) $S_{46}$ of $\beta$-Dipeptide ({\bf 9}).}
    \label{fig:convergence}
\end{figure*}

\subsection{Single bond dissociation}
Due to the inclusion of both HF and doubly excited configurations in DCIS, it is expected to provide reasonable descriptions of single bond breaking, which is challenging for many existing methods. This is indeed the case: Fig. \ref{fig:DCIS_PES} presents the potential energy curves for (a) hydrogen fluoride and (b) diatomic fluorine molecules computed with a 6-31G basis, along with the energy errors from full CI (FCI) shown in (c) and (d). DCIS successfully yields reasonable potential energy curves.
Notably, the DCIS ground state energy is consistently lower than that of HF, but its correlation effect is limited near the equilibrium bond length, highlighting that the method is size-intensive and primarily captures static correlation. The de-excitation of a CIS state simply reverts to the HF state when static correlation is absent. 
Importantly, these findings suggest that CIS states can serve as reasonably accurate references even when the ground state HF state fails at the dissociation limit. 

In Table \ref{tab:NPE}, the non-parallelity errors (NPEs) against FCI potential curves are summarized. Interestingly, DCIS achieves NPEs of 14.7 mH and 15.9 mH for HF and F$_2$, respectively, which are comparable to---or even smaller than---those of standard CASSCF with a 2-electron-2-orbital active space (20.3 and 24.6 mH). However, we consider this may result from error cancellation in DCIS. 
While CASSCF optimizes orbitals self-consistently and generally yields shallow potential curves, DCIS lacks orbital optimization, resulting in higher energy than CASSCF at the dissociation limit.  Overall, DCIS cancels out the effects of dynamical correlation and orbital optimization, leading to relatively balanced NPEs.

Fig. \ref{fig:DCIS_PES} also depicts the results of SFCIS, another CIS variant known for enabling single-bond breaking in a size-intensive manner. Since SFCIS employs a high-spin triplet state as a reference, the orbitals are well-suited for the dissociation limit but not necessarily for the singlet ground state near equilibrium. Consequently, SFCIS energies tend to be relatively lower at dissociation, resulting in considerably shallow potential curves. {The corresponding} NPEs are 81.7 mH and 38.4 mH. 

Moreover, SFCIS is often contaminated by triplet states especially for excited states, although the ground state  typically remains close to singlet and its spin contamination can be eliminated in several ways.\cite{Tsuchimochi2015, Horbatenko2021} In contrast, DCIS can be made always spin pure, if the underlying HF state is spin-restricted, as in the present examples.

Finally, it should be stressed that both DCIS and SFCIS rely on initial states, namely, the target CIS state and the reference triplet state. It is therefore essential for these methods to include the anti-bonding orbital to describe bond dissociation\cite{Krylov2001}; in the present case, we used the CIS state with a strong $\sigma\rightarrow \sigma^*$ transition.

\subsection{Convergence behavior: maximum-overlap v.s. block Davidson}
Many of the excited states discussed in Section \ref{sec:CT} are high-lying, and can be accessed from the bottom using the Davidson algorithm in CIS. As described in Section \ref{sec:algorithm}, the DCIS solutions also generally appear as higher roots, presenting a similar challenge. 

In this section, we examine the convergence behavior of DCIS using the maximum-overlap Davidson algorithm, comparing it to the block Davidson method across various types of excitations. For this analysis, we selected three systems from the test set used above:  one excited state of benzonitrile ({\bf 4}) and two excited states of $\beta$-Dipeptide ({\bf 9}). 
The third CIS excited state of  benzonitrile is relatively valence in character, with the CIS orbital gradient being $|^{\rm o}{\bf g}| = 0.106$ and a dipole change of $|\Delta {\bm\mu}|=1.01$. The two excited states of $\beta$-Dipeptide corresponding to the 13th and 46th CIS eigenstates both exhibit large gradients and dipole changes, $|^{\rm o}{\bf g}| =0.257$ and $0.290$ and $|\Delta {\bm\mu}|=2.60$ and $4.65$, respectively. 

In the maximum-overlap Davidson algorithm, the initial trial vector is set to the CI coefficients of the target CIS state $I$, i.e., $\bar{\bf d} = {\bf c}^I$ and ${\bf d} = {\bf 0}$. For the block Davidson algorithm, $(I+1)$ initial states (including the HF state) are prepared, with trial vectors chosen as the CI coefficients of the target state and all lower-lying CIS states. The convergence criterion is set to ensure the DCIS gradient norm is smaller than $10^{-5}$.

The results are shown in Fig. \ref{fig:convergence}, which plots the gradient norm of each solution as a function of the number of trial vectors  (i.e., subspace dimension). The block Davidson algorithm handles multiple solutions at each iteration, and often provides stable convergence for all states simultaneously. In practice, the number of cycles required to converge all states is 10-15, regardless of the system. However, a major drawback is the rapid expansion of the subspace; namely, at each iteration, the dimension increases by $I+1$ additional trial vectors. As a result, the computational cost required to converge the target state becomes increasingly larger for higher-lying CIS states. For example, while the $S_3$ state of {\bf 4} (Fig. \ref{fig:convergence}(a)) requires a subspace with 41 vectors to converge the target state (red curve), the $S_{46}$ state of {\bf 9} (Fig. \ref{fig:convergence}(c)) requires 396 vectors (the number of cycles before convergence is 13 and 10, respectively). 

In contrast, it is clearly shown in Fig. \ref{fig:convergence} that the maximum-overlap algorithm (green curve) converges with a much smaller subspace dimension compared to the block Davidson algorithm. For the $S_3$ state of {\bf 4}, it  takes only 14 cycles (vectors), achieving a threefold  cost reduction compared to the block Davidson method. However, when the target state has a larger $|^{\rm o}{\bf g}|$, as in the two excitations of {\bf 9}, it generally requires more iterations and vectors to converge, due to the larger relaxation effect. Specifically, the maximum-overlap method converges with 30 and 45 vectors for $S_{13}$ and $S_{46}$ of ({\bf 9}), respectively. At some iterations, the DCIS gradient increases compared to the previous iteration, typically indicating the emergence of intruder (low-lying) state in the subspace. Nonetheless, convergence is still achieved stably, and the sizes of these subspaces are 4.2 and 8.8 times smaller than those of the block Davidson method, offering a significant reduction in cost.

\section{Concluding remarks}
In this work, we first derived the CIS electronic Hessian, and building upon this foundation, introduced a new size-intensive method, Double CIS. DCIS approximately handles the orbital relaxation of the target CIS state to first order. Thanks to the variational nature of DCIS, it can be expressed as a generalized eigenvalue problem, effectively representing ``CIS on top of CIS,'' which enables stable optimization in contrast to the challenges associated with the non-linear full orbital optimization. Furthermore, we proposed an efficient approach for the direct optimization in DCIS, particularly useful for high-lying excited states, by evading the determination of lower-lying states. 
Our results show significant improvements in the descriptions of charge transfer excitations and bond breaking systems. We anticipate that incorporating dynamical correlation will further enhance these results, which we plan to address in future work.

\section*{Acknowledgement}
This work was supported by the start-up funding from the Shibaura Institute of Technology and JST FOREST Program, Grant No. JPMJFR223U.

\section*{Data Availability}
The data that supports the findings of this study are available within the article and its supplementary material.

\section*{Supplementary Material}
The Supplementary Material includes: (1) a detailed derivation of the analytical expression for the CIS Hessian, (2) a discussion on the equivalence of different DCIS parameterizations, (3) the pseudocode for the maximum-overlap Davidson algorithm, (4) a table listing excitation energies, and (5) tables presenting the total energies of the HF and F$_2$ molecules.

\bibliographystyle{apsrev4-1}
\providecommand{\noopsort}[1]{}\providecommand{\singleletter}[1]{#1}%

\end{document}

%% file: settings.tex
\usepackage{graphicx}
\usepackage{amsmath,amsfonts,amssymb}
\usepackage{color}
\usepackage{bm}
\usepackage[mathscr]{eucal}
\usepackage{picture}
\usepackage{float}
\usepackage{subcaption}

\usepackage{graphicx}
\usepackage{amsmath,amsfonts,amssymb}
\usepackage{color}
\usepackage{bm}
\usepackage{threeparttable}
\usepackage[mathscr]{eucal}
\usepackage{picture}

\definecolor{grey}{rgb}{0.7,0.7,0.7}

\newcommand{\br}{\nonumber\\}
\usepackage{fancyvrb}
\usepackage{ulem}

\usepackage[T1]{fontenc}

\definecolor{brown}{RGB}{111,16,50}
\definecolor{purple}{rgb}{0.8,0.0,0.8}
\definecolor{pink}{rgb}{1.,0.5,0.5}

\newcommand{\ctext}[1]{\raise0.2ex\hbox{\textcircled{\scriptsize{#1}}}}

%% file: resub.bbl
\begin{thebibliography}{37}%
\makeatletter
\providecommand \@ifxundefined [1]{%
 \@ifx{#1\undefined}
}%
\providecommand \@ifnum [1]{%
 \ifnum #1\expandafter \@firstoftwo
 \else \expandafter \@secondoftwo
 \fi
}%
\providecommand \@ifx [1]{%
 \ifx #1\expandafter \@firstoftwo
 \else \expandafter \@secondoftwo
 \fi
}%
\providecommand \natexlab [1]{#1}%
\providecommand \enquote  [1]{``#1''}%
\providecommand \bibnamefont  [1]{#1}%
\providecommand \bibfnamefont [1]{#1}%
\providecommand \citenamefont [1]{#1}%
\providecommand \href@noop [0]{\@secondoftwo}%
\providecommand \href [0]{\begingroup \@sanitize@url \@href}%
\providecommand \@href[1]{\@@startlink{#1}\@@href}%
\providecommand \@@href[1]{\endgroup#1\@@endlink}%
\providecommand \@sanitize@url [0]{\catcode `\\12\catcode `\$12\catcode
  `\&12\catcode `\#12\catcode `\^12\catcode `\_12\catcode `\%12\relax}%
\providecommand \@@startlink[1]{}%
\providecommand \@@endlink[0]{}%
\providecommand \url  [0]{\begingroup\@sanitize@url \@url }%
\providecommand \@url [1]{\endgroup\@href {#1}{\urlprefix }}%
\providecommand \urlprefix  [0]{URL }%
\providecommand \Eprint [0]{\href }%
\providecommand \doibase [0]{http://dx.doi.org/}%
\providecommand \selectlanguage [0]{\@gobble}%
\providecommand \bibinfo  [0]{\@secondoftwo}%
\providecommand \bibfield  [0]{\@secondoftwo}%
\providecommand \translation [1]{[#1]}%
\providecommand \BibitemOpen [0]{}%
\providecommand \bibitemStop [0]{}%
\providecommand \bibitemNoStop [0]{.\EOS\space}%
\providecommand \EOS [0]{\spacefactor3000\relax}%
\providecommand \BibitemShut  [1]{\csname bibitem#1\endcsname}%
\let\auto@bib@innerbib\@empty
%</preamble>
\bibitem [{\citenamefont {Kowalczyk}\ \emph {et~al.}(2011)\citenamefont
  {Kowalczyk}, \citenamefont {Yost},\ and\ \citenamefont
  {Voorhis}}]{Kowalczyk2011}%
  \BibitemOpen
  \bibfield  {author} {\bibinfo {author} {\bibfnamefont {T.}~\bibnamefont
  {Kowalczyk}}, \bibinfo {author} {\bibfnamefont {S.~R.}\ \bibnamefont {Yost}},
  \ and\ \bibinfo {author} {\bibfnamefont {T.~V.}\ \bibnamefont {Voorhis}},\
  }\href {\doibase 10.1063/1.3530801} {\bibfield  {journal} {\bibinfo
  {journal} {J. Chem. Phys.}\ }\textbf {\bibinfo {volume} {134}},\ \bibinfo
  {pages} {054128} (\bibinfo {year} {2011})}\BibitemShut {NoStop}%
\bibitem [{\citenamefont {Kowalczyk}\ \emph {et~al.}(2013)\citenamefont
  {Kowalczyk}, \citenamefont {Tsuchimochi}, \citenamefont {Chen}, \citenamefont
  {Top},\ and\ \citenamefont {Van~Voorhis}}]{Kowalczyk2013}%
  \BibitemOpen
  \bibfield  {author} {\bibinfo {author} {\bibfnamefont {T.}~\bibnamefont
  {Kowalczyk}}, \bibinfo {author} {\bibfnamefont {T.}~\bibnamefont
  {Tsuchimochi}}, \bibinfo {author} {\bibfnamefont {P.-T.}\ \bibnamefont
  {Chen}}, \bibinfo {author} {\bibfnamefont {L.}~\bibnamefont {Top}}, \ and\
  \bibinfo {author} {\bibfnamefont {T.}~\bibnamefont {Van~Voorhis}},\ }\href
  {\doibase 10.1063/1.4801790} {\bibfield  {journal} {\bibinfo  {journal} {J.
  Chem. Phys.}\ }\textbf {\bibinfo {volume} {138}},\ \bibinfo {pages} {164101}
  (\bibinfo {year} {2013})}
  \BibitemShut {NoStop}%
\bibitem [{\citenamefont {Levi}\ \emph {et~al.}(2020)\citenamefont {Levi},
  \citenamefont {Ivanov},\ and\ \citenamefont {Jónsson}}]{Levi2020}%
  \BibitemOpen
  \bibfield  {author} {\bibinfo {author} {\bibfnamefont {G.}~\bibnamefont
  {Levi}}, \bibinfo {author} {\bibfnamefont {A.~V.}\ \bibnamefont {Ivanov}}, \
  and\ \bibinfo {author} {\bibfnamefont {H.}~\bibnamefont {Jónsson}},\ }\href
  {\doibase 10.1021/acs.jctc.0c00597} {\bibfield  {journal} {\bibinfo
  {journal} {J. Chem. Theory Comput.}\ }\textbf {\bibinfo {volume} {16}},\
  \bibinfo {pages} {6968} (\bibinfo {year} {2020})}\BibitemShut {NoStop}%
\bibitem [{\citenamefont {Hait}\ and\ \citenamefont
  {Head-Gordon}(2021)}]{Hait2021}%
  \BibitemOpen
  \bibfield  {author} {\bibinfo {author} {\bibfnamefont {D.}~\bibnamefont
  {Hait}}\ and\ \bibinfo {author} {\bibfnamefont {M.}~\bibnamefont
  {Head-Gordon}},\ }\href {\doibase 10.1021/acs.jpclett.1c00744} {\bibfield
  {journal} {\bibinfo  {journal} {J. Phys. Chem. Lett.}\ }\textbf {\bibinfo
  {volume} {12}},\ \bibinfo {pages} {4517} (\bibinfo {year}
  {2021})}\BibitemShut {NoStop}%
\bibitem [{\citenamefont {Burton}(2022)}]{Burton2022}%
  \BibitemOpen
  \bibfield  {author} {\bibinfo {author} {\bibfnamefont {H.~G.}\ \bibnamefont
  {Burton}},\ }\href {\doibase 10.1021/acs.jctc.1c01089} {\bibfield  {journal}
  {\bibinfo  {journal} {J. Chem. Theory Comput.}\ }\textbf {\bibinfo {volume}
  {18}},\ \bibinfo {pages} {1512} (\bibinfo {year} {2022})}\BibitemShut
  {NoStop}%
\bibitem [{\citenamefont {Kossoski}\ and\ \citenamefont
  {Loos}(2023)}]{Kossoski2023}%
  \BibitemOpen
  \bibfield  {author} {\bibinfo {author} {\bibfnamefont {F.}~\bibnamefont
  {Kossoski}}\ and\ \bibinfo {author} {\bibfnamefont {P.-F.}\ \bibnamefont
  {Loos}},\ }\href {\doibase 10.1021/acs.jctc.3c00057} {\bibfield  {journal}
  {\bibinfo  {journal} {J. Chem. Theory Comput.}\ }\textbf {\bibinfo {volume}
  {19}},\ \bibinfo {pages} {2258} (\bibinfo {year} {2023})}\BibitemShut
  {NoStop}%
\bibitem [{\citenamefont {Marie}\ and\ \citenamefont
  {Burton}(2023)}]{Marie2023}%
  \BibitemOpen
  \bibfield  {author} {\bibinfo {author} {\bibfnamefont {A.}~\bibnamefont
  {Marie}}\ and\ \bibinfo {author} {\bibfnamefont {H.~G.~A.}\ \bibnamefont
  {Burton}},\ }\href {\doibase 10.1021/acs.jpca.3c00603} {\bibfield  {journal}
  {\bibinfo  {journal} {JPCA}\ }\textbf {\bibinfo {volume} {127}},\ \bibinfo
  {pages} {4538} (\bibinfo {year} {2023})}\BibitemShut {NoStop}%
\bibitem [{\citenamefont {Tuckman}\ and\ \citenamefont
  {Neuscamman}(2023)}]{Tuckman2023}%
  \BibitemOpen
  \bibfield  {author} {\bibinfo {author} {\bibfnamefont {H.}~\bibnamefont
  {Tuckman}}\ and\ \bibinfo {author} {\bibfnamefont {E.}~\bibnamefont
  {Neuscamman}},\ }\href {\doibase 10.1021/acs.jctc.3c00194} {\bibfield
  {journal} {\bibinfo  {journal} {J. Chem. Theory Comput.}\ }\textbf {\bibinfo
  {volume} {19}},\ \bibinfo {pages} {6160} (\bibinfo {year}
  {2023})}\BibitemShut {NoStop}%
\bibitem [{\citenamefont {Damour}\ \emph {et~al.}(2024)\citenamefont {Damour},
  \citenamefont {Scemama}, \citenamefont {Jacquemin}, \citenamefont
  {Kossoski},\ and\ \citenamefont {Loos}}]{Damour2024}%
  \BibitemOpen
  \bibfield  {author} {\bibinfo {author} {\bibfnamefont {Y.}~\bibnamefont
  {Damour}}, \bibinfo {author} {\bibfnamefont {A.}~\bibnamefont {Scemama}},
  \bibinfo {author} {\bibfnamefont {D.}~\bibnamefont {Jacquemin}}, \bibinfo
  {author} {\bibfnamefont {F.}~\bibnamefont {Kossoski}}, \ and\ \bibinfo
  {author} {\bibfnamefont {P.-F.}\ \bibnamefont {Loos}},\ }\href {\doibase
  10.1021/acs.jctc.4c00034} {\bibfield  {journal} {\bibinfo  {journal} {J.
  Chem. Theory Comput.}\ }\textbf {\bibinfo {volume} {20}},\ \bibinfo {pages}
  {4129} (\bibinfo {year} {2024})}\BibitemShut {NoStop}%
\bibitem [{\citenamefont {Selenius}\ \emph {et~al.}(2024)\citenamefont
  {Selenius}, \citenamefont {Sigurdarson}, \citenamefont {Schmerwitz},\ and\
  \citenamefont {Levi}}]{Selenius2024}%
  \BibitemOpen
  \bibfield  {author} {\bibinfo {author} {\bibfnamefont {E.}~\bibnamefont
  {Selenius}}, \bibinfo {author} {\bibfnamefont {A.~E.}\ \bibnamefont
  {Sigurdarson}}, \bibinfo {author} {\bibfnamefont {Y.~L.~A.}\ \bibnamefont
  {Schmerwitz}}, \ and\ \bibinfo {author} {\bibfnamefont {G.}~\bibnamefont
  {Levi}},\ }\href {\doibase 10.1021/acs.jctc.3c01319} {\bibfield  {journal}
  {\bibinfo  {journal} {J. Chem. Theory Comput.}\ }\textbf {\bibinfo {volume}
  {20}},\ \bibinfo {pages} {3809} (\bibinfo {year} {2024})}\BibitemShut
  {NoStop}%
\bibitem [{\citenamefont {Hait}\ and\ \citenamefont
  {Head-Gordon}(2020{\natexlab{a}})}]{Hait2020A}%
  \BibitemOpen
  \bibfield  {author} {\bibinfo {author} {\bibfnamefont {D.}~\bibnamefont
  {Hait}}\ and\ \bibinfo {author} {\bibfnamefont {M.}~\bibnamefont
  {Head-Gordon}},\ }\href {\doibase 10.1021/acs.jpclett.9b03661} {\bibfield
  {journal} {\bibinfo  {journal} {J. Phys. Chem. Lett.}\ }\textbf {\bibinfo
  {volume} {11}},\ \bibinfo {pages} {775} (\bibinfo {year}
  {2020}{\natexlab{a}})},\ \bibinfo {note} {pMID: 31917579}\BibitemShut
  {NoStop}%
\bibitem [{\citenamefont {Garner}\ and\ \citenamefont
  {Neuscamman}(2020)}]{Garner2020}%
  \BibitemOpen
  \bibfield  {author} {\bibinfo {author} {\bibfnamefont {S.~M.}\ \bibnamefont
  {Garner}}\ and\ \bibinfo {author} {\bibfnamefont {E.}~\bibnamefont
  {Neuscamman}},\ }\href {\doibase 10.1063/5.0020595} {\bibfield  {journal}
  {\bibinfo  {journal} {J. Chem. Phys.}\ }\textbf {\bibinfo {volume} {153}},\
  \bibinfo {pages} {154102} (\bibinfo {year} {2020})}\BibitemShut {NoStop}%
\bibitem [{\citenamefont {Cunha}\ \emph {et~al.}(2022)\citenamefont {Cunha},
  \citenamefont {Hait}, \citenamefont {Kang}, \citenamefont {Mao},\ and\
  \citenamefont {Head-Gordon}}]{Cunha2022}%
  \BibitemOpen
  \bibfield  {author} {\bibinfo {author} {\bibfnamefont {L.~A.}\ \bibnamefont
  {Cunha}}, \bibinfo {author} {\bibfnamefont {D.}~\bibnamefont {Hait}},
  \bibinfo {author} {\bibfnamefont {R.}~\bibnamefont {Kang}}, \bibinfo {author}
  {\bibfnamefont {Y.}~\bibnamefont {Mao}}, \ and\ \bibinfo {author}
  {\bibfnamefont {M.}~\bibnamefont {Head-Gordon}},\ }\href {\doibase
  10.1021/acs.jpclett.2c00578} {\bibfield  {journal} {\bibinfo  {journal} {J.
  Phys. Chem. Lett.}\ }\textbf {\bibinfo {volume} {13}},\ \bibinfo {pages}
  {3438} (\bibinfo {year} {2022})}\BibitemShut {NoStop}%
\bibitem [{\citenamefont {Arias-Martinez}\ \emph {et~al.}(2022)\citenamefont
  {Arias-Martinez}, \citenamefont {Cunha}, \citenamefont {Oosterbaan},
  \citenamefont {Lee},\ and\ \citenamefont {Head-Gordon}}]{arias-martinez2022}%
  \BibitemOpen
  \bibfield  {author} {\bibinfo {author} {\bibfnamefont {J.~E.}\ \bibnamefont
  {Arias-Martinez}}, \bibinfo {author} {\bibfnamefont {L.~A.}\ \bibnamefont
  {Cunha}}, \bibinfo {author} {\bibfnamefont {K.~J.}\ \bibnamefont
  {Oosterbaan}}, \bibinfo {author} {\bibfnamefont {J.}~\bibnamefont {Lee}}, \
  and\ \bibinfo {author} {\bibfnamefont {M.}~\bibnamefont {Head-Gordon}},\
  }\href@noop {} {\bibfield  {journal} {\bibinfo  {journal} {Phys. Chem. Chem.
  Phys.}\ }\textbf {\bibinfo {volume} {24}},\ \bibinfo {pages} {20728}
  (\bibinfo {year} {2022})}\BibitemShut {NoStop}%
\bibitem [{\citenamefont {Liu}\ \emph {et~al.}(2012)\citenamefont {Liu},
  \citenamefont {Fatehi}, \citenamefont {Shao}, \citenamefont {Veldkamp},\ and\
  \citenamefont {Subotnik}}]{Liu2012}%
  \BibitemOpen
  \bibfield  {author} {\bibinfo {author} {\bibfnamefont {X.}~\bibnamefont
  {Liu}}, \bibinfo {author} {\bibfnamefont {S.}~\bibnamefont {Fatehi}},
  \bibinfo {author} {\bibfnamefont {Y.}~\bibnamefont {Shao}}, \bibinfo {author}
  {\bibfnamefont {B.~S.}\ \bibnamefont {Veldkamp}}, \ and\ \bibinfo {author}
  {\bibfnamefont {J.~E.}\ \bibnamefont {Subotnik}},\ }\href {\doibase
  10.1063/1.4705757} {\bibfield  {journal} {\bibinfo  {journal} {J. Chem.
  Phys.}\ }\textbf {\bibinfo {volume} {136}},\ \bibinfo {pages} {161101}
  (\bibinfo {year} {2012})}\BibitemShut {NoStop}%
\bibitem [{\citenamefont {Meyer}\ \emph {et~al.}(2014)\citenamefont {Meyer},
  \citenamefont {Domingo}, \citenamefont {Krah},\ and\ \citenamefont
  {Robert}}]{Meyer2014}%
  \BibitemOpen
  \bibfield  {author} {\bibinfo {author} {\bibfnamefont {B.}~\bibnamefont
  {Meyer}}, \bibinfo {author} {\bibfnamefont {A.}~\bibnamefont {Domingo}},
  \bibinfo {author} {\bibfnamefont {T.}~\bibnamefont {Krah}}, \ and\ \bibinfo
  {author} {\bibfnamefont {V.}~\bibnamefont {Robert}},\ }\href {\doibase
  10.1039/C4DT00768H} {\bibfield  {journal} {\bibinfo  {journal} {Dalton
  Trans.}\ }\textbf {\bibinfo {volume} {43}},\ \bibinfo {pages} {11209}
  (\bibinfo {year} {2014})}\BibitemShut {NoStop}%
\bibitem [{\citenamefont {Hait}\ \emph {et~al.}(2016)\citenamefont {Hait},
  \citenamefont {Zhu}, \citenamefont {McMahon},\ and\ \citenamefont
  {Van~Voorhis}}]{Hait2016}%
  \BibitemOpen
  \bibfield  {author} {\bibinfo {author} {\bibfnamefont {D.}~\bibnamefont
  {Hait}}, \bibinfo {author} {\bibfnamefont {T.}~\bibnamefont {Zhu}}, \bibinfo
  {author} {\bibfnamefont {D.~P.}\ \bibnamefont {McMahon}}, \ and\ \bibinfo
  {author} {\bibfnamefont {T.}~\bibnamefont {Van~Voorhis}},\ }\href {\doibase
  10.1021/acs.jctc.6b00426} {\bibfield  {journal} {\bibinfo  {journal} {J.
  Chem. Theory Comput.}\ }\textbf {\bibinfo {volume} {12}},\ \bibinfo {pages}
  {3353} (\bibinfo {year} {2016})}\BibitemShut {NoStop}%
\bibitem [{\citenamefont {Zhao}\ and\ \citenamefont
  {Neuscamman}(2020)}]{Zhao2020}%
  \BibitemOpen
  \bibfield  {author} {\bibinfo {author} {\bibfnamefont {L.}~\bibnamefont
  {Zhao}}\ and\ \bibinfo {author} {\bibfnamefont {E.}~\bibnamefont
  {Neuscamman}},\ }\href {\doibase 10.1021/acs.jctc.9b00530} {\bibfield
  {journal} {\bibinfo  {journal} {J. Chem. Theory Comput.}\ }\textbf {\bibinfo
  {volume} {16}},\ \bibinfo {pages} {164} (\bibinfo {year} {2020})}\BibitemShut
  {NoStop}%
\bibitem [{\citenamefont {Subotnik}(2011)}]{Subotnik2011}%
  \BibitemOpen
  \bibfield  {author} {\bibinfo {author} {\bibfnamefont {J.~E.}\ \bibnamefont
  {Subotnik}},\ }\href {\doibase 10.1063/1.3627152} {\bibfield  {journal}
  {\bibinfo  {journal} {J. Chem. Phys.}\ }\textbf {\bibinfo {volume} {135}},\
  \bibinfo {pages} {071104} (\bibinfo {year} {2011})}\BibitemShut {NoStop}%
\bibitem [{\citenamefont {Shea}\ and\ \citenamefont
  {Neuscamman}(2018)}]{Shea2018}%
  \BibitemOpen
  \bibfield  {author} {\bibinfo {author} {\bibfnamefont {J.~A.~R.}\
  \bibnamefont {Shea}}\ and\ \bibinfo {author} {\bibfnamefont {E.}~\bibnamefont
  {Neuscamman}},\ }\href {\doibase 10.1063/1.5045056} {\bibfield  {journal}
  {\bibinfo  {journal} {J. Chem. Phys.}\ }\textbf {\bibinfo {volume} {149}},\
  \bibinfo {pages} {081101} (\bibinfo {year} {2018})}\BibitemShut {NoStop}%
\bibitem [{\citenamefont {Shea}\ \emph {et~al.}(2020)\citenamefont {Shea},
  \citenamefont {Gwin},\ and\ \citenamefont {Neuscamman}}]{Shea2020}%
  \BibitemOpen
  \bibfield  {author} {\bibinfo {author} {\bibfnamefont {J.~A.~R.}\
  \bibnamefont {Shea}}, \bibinfo {author} {\bibfnamefont {E.}~\bibnamefont
  {Gwin}}, \ and\ \bibinfo {author} {\bibfnamefont {E.}~\bibnamefont
  {Neuscamman}},\ }\href {\doibase 10.1021/acs.jctc.9b01105} {\bibfield
  {journal} {\bibinfo  {journal} {J. Chem. Theory Comput.}\ }\textbf {\bibinfo
  {volume} {16}},\ \bibinfo {pages} {1526} (\bibinfo {year}
  {2020})}\BibitemShut {NoStop}%
\bibitem [{\citenamefont {Hardikar}\ and\ \citenamefont
  {Neuscamman}(2020)}]{Hardikar2020}%
  \BibitemOpen
  \bibfield  {author} {\bibinfo {author} {\bibfnamefont {T.~S.}\ \bibnamefont
  {Hardikar}}\ and\ \bibinfo {author} {\bibfnamefont {E.}~\bibnamefont
  {Neuscamman}},\ }\href {\doibase 10.1063/5.0019557} {\bibfield  {journal}
  {\bibinfo  {journal} {J. Chem. Phys.}\ }\textbf {\bibinfo {volume} {153}},\
  \bibinfo {pages} {164108} (\bibinfo {year} {2020})}\BibitemShut {NoStop}%
\bibitem [{\citenamefont {Liu}\ \emph {et~al.}(2013)\citenamefont {Liu},
  \citenamefont {Ou}, \citenamefont {Alguire},\ and\ \citenamefont
  {Subotnik}}]{Liu2013}%
  \BibitemOpen
  \bibfield  {author} {\bibinfo {author} {\bibfnamefont {X.}~\bibnamefont
  {Liu}}, \bibinfo {author} {\bibfnamefont {Q.}~\bibnamefont {Ou}}, \bibinfo
  {author} {\bibfnamefont {E.}~\bibnamefont {Alguire}}, \ and\ \bibinfo
  {author} {\bibfnamefont {J.~E.}\ \bibnamefont {Subotnik}},\ }\href {\doibase
  10.1063/1.4809571} {\bibfield  {journal} {\bibinfo  {journal} {J. Chem.
  Phys.}\ }\textbf {\bibinfo {volume} {138}},\ \bibinfo {pages} {221105}
  (\bibinfo {year} {2013})}\BibitemShut {NoStop}%
\bibitem [{\citenamefont {Helgaker}\ \emph {et~al.}(2000)\citenamefont
  {Helgaker}, \citenamefont {Jørgensen},\ and\ \citenamefont
  {Olsen}}]{Helgaker2000}%
  \BibitemOpen
  \bibfield  {author} {\bibinfo {author} {\bibfnamefont {T.}~\bibnamefont
  {Helgaker}}, \bibinfo {author} {\bibfnamefont {P.}~\bibnamefont
  {Jørgensen}}, \ and\ \bibinfo {author} {\bibfnamefont {J.}~\bibnamefont
  {Olsen}},\ }\href@noop {} {\emph {\bibinfo {title} {Molecular
  Electronic-Structure Theory}}}\ (\bibinfo  {publisher} {Wiley},\ \bibinfo
  {address} {New York},\ \bibinfo {year} {2000})\ p.\ \bibinfo {pages}
  {938}\BibitemShut {NoStop}%
\bibitem [{si()}]{si}%
  \BibitemOpen
  \href@noop {} {}\bibinfo {note} {Supplementary material is
  available.}\BibitemShut {Stop}%
\bibitem [{\citenamefont {Gilbert}\ \emph {et~al.}(2008)\citenamefont
  {Gilbert}, \citenamefont {Besley},\ and\ \citenamefont {Gill}}]{Gilbert2008}%
  \BibitemOpen
  \bibfield  {author} {\bibinfo {author} {\bibfnamefont {A.~T.}\ \bibnamefont
  {Gilbert}}, \bibinfo {author} {\bibfnamefont {N.~A.}\ \bibnamefont {Besley}},
  \ and\ \bibinfo {author} {\bibfnamefont {P.~M.}\ \bibnamefont {Gill}},\
  }\href {\doibase 10.1021/jp801738f} {\bibfield  {journal} {\bibinfo
  {journal} {J. Phys. Chem. A}\ }\textbf {\bibinfo {volume} {112}},\ \bibinfo
  {pages} {13164} (\bibinfo {year} {2008})}\BibitemShut {NoStop}%
\bibitem [{\citenamefont {Hait}\ and\ \citenamefont
  {Head-Gordon}(2020{\natexlab{b}})}]{Hait2020B}%
  \BibitemOpen
  \bibfield  {author} {\bibinfo {author} {\bibfnamefont {D.}~\bibnamefont
  {Hait}}\ and\ \bibinfo {author} {\bibfnamefont {M.}~\bibnamefont
  {Head-Gordon}},\ }\href {\doibase 10.1021/acs.jctc.9b01127} {\bibfield
  {journal} {\bibinfo  {journal} {J. Chem. Theory Comput.}\ }\textbf {\bibinfo
  {volume} {16}},\ \bibinfo {pages} {1699} (\bibinfo {year}
  {2020}{\natexlab{b}})}\BibitemShut {NoStop}%
\bibitem [{\citenamefont {Carter-Fenk}\ and\ \citenamefont
  {Herbert}(2020)}]{CarterFenk2020}%
  \BibitemOpen
  \bibfield  {author} {\bibinfo {author} {\bibfnamefont {K.}~\bibnamefont
  {Carter-Fenk}}\ and\ \bibinfo {author} {\bibfnamefont {J.~M.}\ \bibnamefont
  {Herbert}},\ }\href {\doibase 10.1021/acs.jctc.0c00502} {\bibfield  {journal}
  {\bibinfo  {journal} {J. Chem. Theory Comput.}\ }\textbf {\bibinfo {volume}
  {16}},\ \bibinfo {pages} {5067} (\bibinfo {year} {2020})}\BibitemShut
  {NoStop}%
\bibitem [{\citenamefont {Liu}\ and\ \citenamefont {Subotnik}(2014)}]{Liu2014}%
  \BibitemOpen
  \bibfield  {author} {\bibinfo {author} {\bibfnamefont {X.}~\bibnamefont
  {Liu}}\ and\ \bibinfo {author} {\bibfnamefont {J.~E.}\ \bibnamefont
  {Subotnik}},\ }\href {\doibase 10.1021/ct4009377} {\bibfield  {journal}
  {\bibinfo  {journal} {J. Chem. Theory Comput.}\ }\textbf {\bibinfo {volume}
  {10}},\ \bibinfo {pages} {1004} (\bibinfo {year} {2014})}\BibitemShut
  {NoStop}%
\bibitem [{\citenamefont {Davidson}(1975)}]{Davidson1975}%
  \BibitemOpen
  \bibfield  {author} {\bibinfo {author} {\bibfnamefont {E.~R.}\ \bibnamefont
  {Davidson}},\ }\href {\doibase 10.1016/0021-9991(75)90065-0} {\bibfield
  {journal} {\bibinfo  {journal} {J. Comput. Phys.}\ }\textbf {\bibinfo
  {volume} {17}},\ \bibinfo {pages} {87} (\bibinfo {year} {1975})}\BibitemShut
  {NoStop}%
\bibitem [{\citenamefont {Sadkane}\ and\ \citenamefont
  {Sidje}(1999)}]{Sadkane1999}%
  \BibitemOpen
  \bibfield  {author} {\bibinfo {author} {\bibfnamefont {M.}~\bibnamefont
  {Sadkane}}\ and\ \bibinfo {author} {\bibfnamefont {R.~B.}\ \bibnamefont
  {Sidje}},\ }\href {\doibase 10.1023/A:1019199700323} {\bibfield  {journal}
  {\bibinfo  {journal} {Numerical Algorithms}\ }\textbf {\bibinfo {volume}
  {20}},\ \bibinfo {pages} {217} (\bibinfo {year} {1999})}\BibitemShut
  {NoStop}%
\bibitem [{\citenamefont {Pulay}(1980)}]{Pulay1980}%
  \BibitemOpen
  \bibfield  {author} {\bibinfo {author} {\bibfnamefont {P.}~\bibnamefont
  {Pulay}},\ }\href@noop {} {\bibfield  {journal} {\bibinfo  {journal} {Chem.
  Phys. Lett.}\ }\textbf {\bibinfo {volume} {73}},\ \bibinfo {pages} {393}
  (\bibinfo {year} {1980})}\BibitemShut {NoStop}%
\bibitem [{\citenamefont {Pulay}(1982)}]{Pulay1982}%
  \BibitemOpen
  \bibfield  {author} {\bibinfo {author} {\bibfnamefont {P.}~\bibnamefont
  {Pulay}},\ }\href@noop {} {\bibfield  {journal} {\bibinfo  {journal} {J.
  Comp. Chem.}\ }\textbf {\bibinfo {volume} {3}},\ \bibinfo {pages} {556}
  (\bibinfo {year} {1982})}\BibitemShut {NoStop}%
\bibitem [{\citenamefont {Loos}\ \emph {et~al.}(2021)\citenamefont {Loos},
  \citenamefont {Comin}, \citenamefont {Blase},\ and\ \citenamefont
  {Jacquemin}}]{Loos2021}%
  \BibitemOpen
  \bibfield  {author} {\bibinfo {author} {\bibfnamefont {P.-F.}\ \bibnamefont
  {Loos}}, \bibinfo {author} {\bibfnamefont {M.}~\bibnamefont {Comin}},
  \bibinfo {author} {\bibfnamefont {X.}~\bibnamefont {Blase}}, \ and\ \bibinfo
  {author} {\bibfnamefont {D.}~\bibnamefont {Jacquemin}},\ }\href {\doibase
  10.1021/acs.jctc.1c00226} {\bibfield  {journal} {\bibinfo  {journal} {J.
  Chem. Theory Comput.}\ }\textbf {\bibinfo {volume} {17}},\ \bibinfo {pages}
  {3671} (\bibinfo {year} {2021})}\BibitemShut {NoStop}%
\bibitem [{\citenamefont {Krylov}(2001)}]{Krylov2001}%
  \BibitemOpen
  \bibfield  {author} {\bibinfo {author} {\bibfnamefont {A.~I.}\ \bibnamefont
  {Krylov}},\ }\href {\doibase https://doi.org/10.1016/S0009-2614(01)00287-1}
  {\bibfield  {journal} {\bibinfo  {journal} {Chem. Phys. Lett.}\ }\textbf
  {\bibinfo {volume} {338}},\ \bibinfo {pages} {375} (\bibinfo {year}
  {2001})}\BibitemShut {NoStop}%
\bibitem [{\citenamefont {Tsuchimochi}(2015)}]{Tsuchimochi2015}%
  \BibitemOpen
  \bibfield  {author} {\bibinfo {author} {\bibfnamefont {T.}~\bibnamefont
  {Tsuchimochi}},\ }\href {\doibase 10.1063/1.4934947} {\bibfield  {journal}
  {\bibinfo  {journal} {J. Chem. Phys.}\ }\textbf {\bibinfo {volume} {143}},\
  \bibinfo {pages} {144114} (\bibinfo {year} {2015})}\BibitemShut {NoStop}%
\bibitem [{\citenamefont {Horbatenko}\ \emph {et~al.}(2021)\citenamefont
  {Horbatenko}, \citenamefont {Sadiq}, \citenamefont {Lee}, \citenamefont
  {Filatov},\ and\ \citenamefont {Choi}}]{Horbatenko2021}%
  \BibitemOpen
  \bibfield  {author} {\bibinfo {author} {\bibfnamefont {Y.}~\bibnamefont
  {Horbatenko}}, \bibinfo {author} {\bibfnamefont {S.}~\bibnamefont {Sadiq}},
  \bibinfo {author} {\bibfnamefont {S.}~\bibnamefont {Lee}}, \bibinfo {author}
  {\bibfnamefont {M.}~\bibnamefont {Filatov}}, \ and\ \bibinfo {author}
  {\bibfnamefont {C.~H.}\ \bibnamefont {Choi}},\ }\href {\doibase
  10.1021/acs.jctc.0c01074} {\bibfield  {journal} {\bibinfo  {journal} {J.
  Chem. Theory Comput.}\ }\textbf {\bibinfo {volume} {17}},\ \bibinfo {pages}
  {848} (\bibinfo {year} {2021})}\BibitemShut {NoStop}%
\end{thebibliography}
